\documentclass[prd,aps,nofootinbib,preprintnumbers]{revtex4}

\usepackage{graphicx}

\begin{document}

\title{ Influence on Starobinsky inflation by other fields with large amplitude}

\author{Shinta Kasuya and Kanto Moritake}

\affiliation{
Department of Mathematics and Physics,
     Kanagawa University, Kanagawa 259-1293, Japan}
     
\date{February 26, 2016}

\begin{abstract}
The Starobinsky model is one of the inflation models consistent with the result of CMB observation by the 
Planck satellite. We consider the dynamics of the Starobinsky inflation in the presence of another scalar field 
with a large expectation value during inflation due to a negative Hubble-induced mass. We find that it would 
be affected if the other field has an amplitude close to the Planck scale. In this case, we may observe such 
effects on the Starobinsky model by future CMB experiments.
\end{abstract}

\maketitle

\section{Introduction} 
Inflation is a very attractive cosmological paradigm. It solves the flatness and homogeneity problems of big 
bang cosmology, and could provide the seeds for the large-scale structure of the Universe as well. 
Inflation is usually considered to be driven by a scalar field called the inflaton which slow-rolls on its potential.
The shape of the potential is carefully chosen in order to have both long enough inflation and density 
perturbations observed. Starobinsky inflation~\cite{Starobinsky} is one of the models consistent with observations 
of the cosmic microwave background by the Planck satellite~\cite{Planck}.

The inflaton potential may, however, be affected by other fields. In Ref.~\cite{chaotic2}, it is considered that the 
potential of the chaotic inflation~\cite{chaotic} could be suppressed by a scalar field with large amplitude due 
to a negative Hubble-induced mass term. In particular, they showed that this effect renders the chaotic inflation 
with quadratic potential consistent with Planck observations, otherwise marginally excluded~\cite{Planck}. It is then 
sensible to think the other way around: Is there any situation in which this mechanism could affect the observationally 
preferable inflation model considerably? Or we may think about any observational evidence for other fields that
have large amplitudes during such inflation in the future CMB experiments.

In this article, we consider the influence on the Starobinsky inflation model by other fields with a large field 
amplitude due to a negative Hubble-induced mass term. As well known, it is reasonable for a scalar field to have 
a negative Hubble-induced mass term in supergravity~\cite{Dine}. Moreover, the Starobinsky inflation is realized 
in supergravity in many ways, such as in no-scale supergravity~\cite{no-scale} or superconformal theory~\cite{superconformal}. 

The structure of the article is as follows:
In the next section, we review the Starobinsky model as a scalar-field-driven inflation. In Sec.~III, we introduce 
another scalar field whose coupling to the inflaton leads to a negative Hubble-induced mass term in the 
Starobinsky model, and we derive numerically the observables such as the spectral index and the tensor-to-scalar
ratio, in addition to the inflaton dynamics in Sec.~IV. In the same section, we obtain the constraints on 
model parameters by current CMB observations and consider the possibility of detecting such effects in future
CMB experiments. In Sec.~V, we present that the model can be  derived naturally in supergravity. We give our 
conclusion in Sec.~VI.
 
\section{$R+R^{2}$ Starobinsky model}
The Starobinsky model~\cite{Starobinsky} is obtained by generalization of Einstein-Hilbert action which 
contains  an $R^{2}$ term as in

\begin{equation}
S= \frac{1}{2} \int d^{4}x \sqrt{-g}  \left( R+\frac{R^{2}}{6M^{2}} \right),
\end{equation}
where $M$ is a mass scale much less than the reduced Planck mass $M_{\rm P}$ (=$2.4 \times 10^{18}$~GeV), 
and $M_{\rm P}=1$ unit is adopted. It is equivalent to canonical gravity with a scalar field $\tilde I$ by conformal 
transformation. Taking the transformation $\bar{g}_{\mu\nu}=(1+ \tilde{I} / 3M^2) g_{\mu \nu}$ and the field 
redefinition $I = \sqrt{\frac{3}{2}} \log{[1+ \tilde{I} / 3M^2]}$, one obtains the redefined action
\begin{equation}
S= \frac{1}{2} \int d^{4}x \sqrt{-\bar{g}} \left[ \bar{R}+(\partial_{\mu} I)^2
- \frac{3}{2} M^{2} \left(1-e^{-\sqrt{\frac{2}{3}} I} \right)^2 \right].
\end{equation}
Thus, the potential of the inflaton $I$ is
\begin{equation}
V_{\rm S} = \frac{3M^{2}}{4} \left (1-e^{- \sqrt{\frac{2}{3}}I} \right )^2.
\end{equation}
The spectral index of the curvature perturbation $n_s$ and the tensor-to-scalar ratio $r$ are described, 
respectively, by slow-roll parameters as $n_s= 1 - 6 \varepsilon_{\rm S} + 2 \eta_{\rm S}$ and 
$r=16 \varepsilon_{\rm S}$, where $\varepsilon_{\rm S}$ and $\eta_{\rm S}$ are given by

\begin{eqnarray}
&& \varepsilon_{\rm S}(I) \equiv \frac{1}{2} \left ( \frac{V'_{\rm S}}{V_{\rm S}} \right )^2 
= \frac{4}{3} \frac{e^{-2 \sqrt{ \frac{2}{3}} I}}{ \left ( 1 - e^{- \sqrt{\frac{2}{3}} I } \right )^{2}},\\
&& \eta_{\rm S}(I) \equiv \frac{V''_{\rm S}}{V_{\rm S}} = - \frac{4}{3} \frac{e^{-\sqrt{ \frac{2}{3}} I} 
\left ( 1 - 2 e^{- \sqrt{\frac{2}{3}} I} \right ) }{ \left ( 1 - e^{- \sqrt{\frac{2}{3}} I } \right ) ^{2}},
\end{eqnarray}
respectively.
The prime denotes a derivative with respect to the inflaton $I$.
The number of $\it e$-folds of slow-roll inflation is calculated as
\begin{equation}
N = \int_{I_{\rm end}}^{I_N} dI \frac{V_{\rm S}}{V'_{\rm S}}= \frac{3}{4} 
\left( e^{\sqrt{\frac{2}{3}} I_N}-e^{\sqrt{\frac{2}{3}} I_{\rm end}} \right) - \frac{\sqrt{6}}{4} \left( I_N - I_{\rm end} \right),
\end{equation}
where $I_N$ and $I_{\rm end}$ represent the amplitudes of the inflaton at the time of $N$ $e$-folds and at the 
end of the inflation [$\eta_{\rm S}(I_{\rm end}) =1$], respectively. In the Starobinsky model, $N$=50 (60) leads to 
$n_s$ = 0.961 (0.968) and $r$ = 0.0042 (0.0030), which are favored by the Planck result~\cite{Planck}.

\section{Starobinsky inflation in the presence of another scalar field with large amplitude}
Let us now introduce another scalar field $\phi$, which obtains large amplitude due to a negative 
Hubble-induced mass term. To illustrate how it may affect the inflaton dynamics, we assume the following 
coupling to the inflaton~\cite{chaotic2}:
\begin{equation}
V(I,\phi) = V_{\rm S}(I) + V(\phi) = V_{\rm S} \left ( 1-c{\phi}^{2} \right ) + \frac{\lambda^{2}}{2q} {\phi}^{2q},
\end{equation}
where $c$ and $\lambda$ are positive coupling constants. As shown later, Eq.~(7) is naturally obtained 
in supergravity. During inflation, $\phi$ stays at the minimum
\begin{equation}
{\phi}_*(I)= \left [ \frac{3c}{2{\lambda^2}}M^2 
\left( 1-e^{-\sqrt{\frac{2}{3}}I} \right)^2 \right]^{\frac{1}{2(q-1)}},
\end{equation}
which changes as $I$ moves during inflation.
Therefore, the inflaton potential receives backreaction to be
\begin{equation}
V(I) = V_{\rm S} \left ( 1-\frac{q-1}{q} c \phi_*^2 \right ),
\end{equation}
where the potential is suppressed at large $I$ compared to $V_{\rm S}$ only.
The slow-roll parameters are thus obtained as
\begin{eqnarray}
&&\varepsilon(I) = \frac{ \left( 1 - c \phi_*^2 \right)^2}{ \left( 1 - \frac{q-1}{q} c \phi_*^2 \right)^2}  \varepsilon_{\rm S}  ,\\
&&\eta(I)= \frac{1 - c \phi_*^2}{1 - \frac{q-1}{q} c \phi_*^2} \eta_{\rm S} 
- \frac{\frac{2}{q-1}c \phi_*^2}{1 - \frac{q-1}{q} c \phi_*^2} \varepsilon_{\rm S}.
\end{eqnarray}
In the next section, we will calculate $I_{N}$ numerically, and estimate the spectral index $n_s(I_N)$ 
and the tensor-to-scalar ratio $r(I_N)$.

\section{spectral index and tensor-to-scalar ratio}
Let us see how the spectral index $n_s$ and the tensor-to-scalar ratio $\it r$ will change in the 
presence of other fields with large amplitude. For given $\lambda$, we use the Planck normalization 
of the primordial density perturbations \cite{Planck},
\begin{equation}
A_{s} = \frac{1}{12 \pi^2} \frac{V^3(I_N)}{V'^2(I_N)} \approx 2.2 \times 10^{-9},
\end{equation}
to determine $M$. 

We show the estimates of $n_s$ and $r$ for various $\lambda$ in Fig. 1, together with $\lambda$ dependences for
$n_s$ and $r$ in Figs. 2 and 3, respectively. We can see that our $n_s$ and $r$ do not differ from those in the pure
Starobinsky model for $\lambda\gtrsim 10^{-4}$. As the amplitude of the $\phi$ field gets closer to the Planck 
scale [$\phi_*(I_N) \gtrsim 0.7~(0.8)$ for $q=3~(8)$, which corresponds to $\lambda \lesssim 10^{-4}$], 
$r$ quickly decreases, while $n_s$ stays in the current Planck-favored region. Lowering $\lambda$ further, 
we have too small $n_s$, which departs from the region allowed by the present Planck observation for 
$\lambda\lesssim 10^{-5}$, i.e., $\phi_*(I_N) \gtrsim 0.99$. Notice that there is little dependence on $q$.

\begin{figure}[h!]
\begin{center}
\begin{tabular}{cc}
\includegraphics[width=80mm]{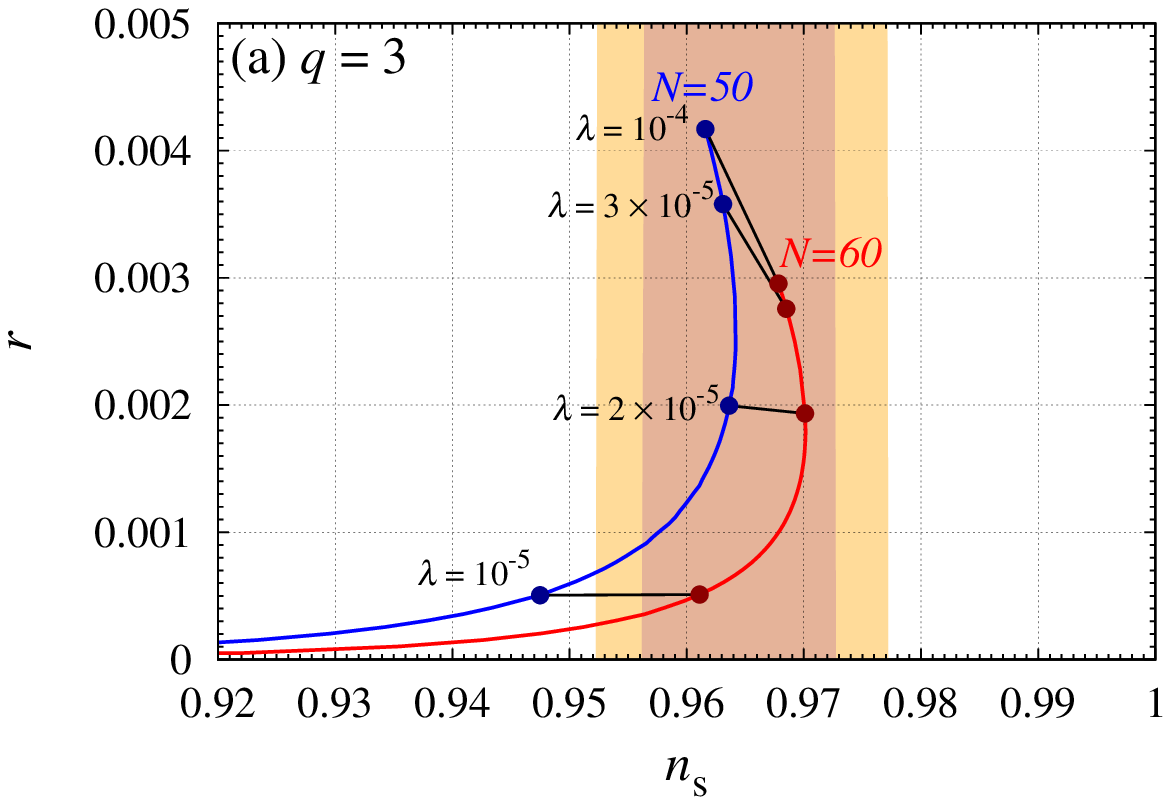} &
\includegraphics[width=80mm]{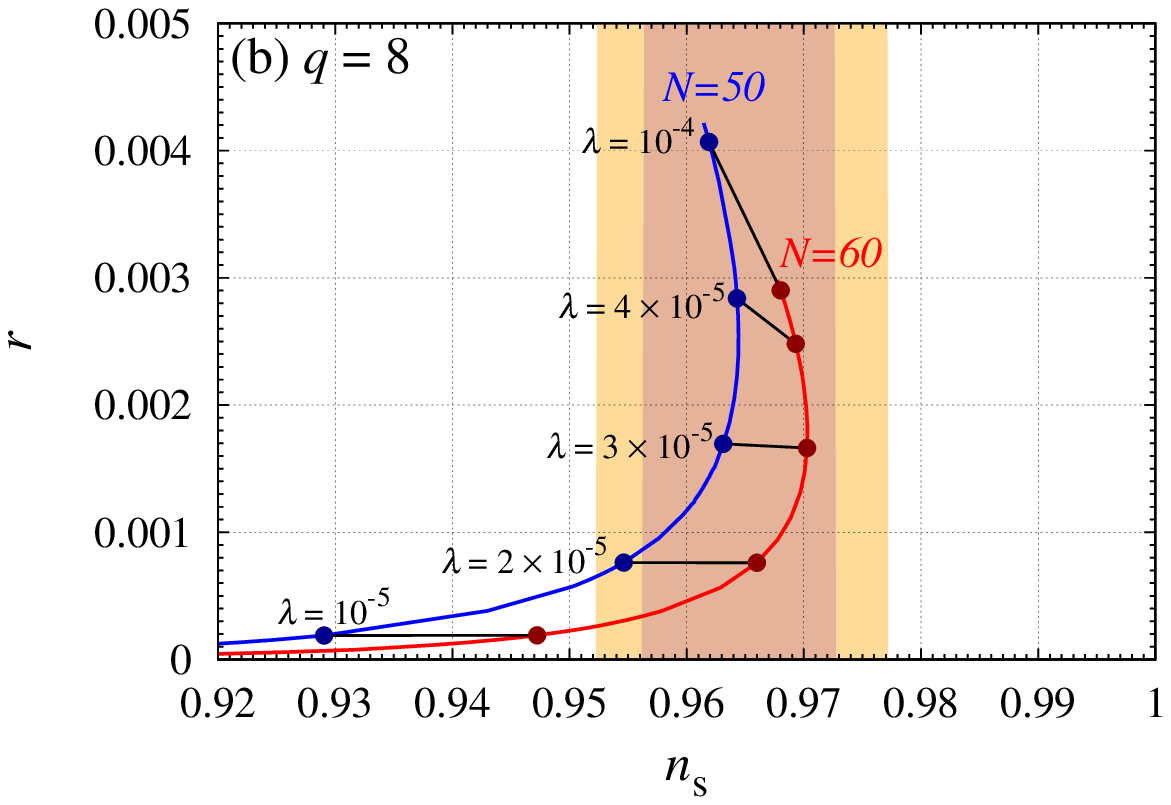}
\end{tabular}
\caption{Spectral index $n_s$ and tensor-to-scalar ratio $r$ for various $\lambda$ for (a) $q=3$ 
and (b) $q=8$. We also show 1$\sigma$ and 2$\sigma$ regions of the Plank result.}
\label{fig1}
\end{center}
\end{figure}

\begin{figure}[h!]
\begin{center}
\begin{tabular}{cc}
\includegraphics[width=80mm]{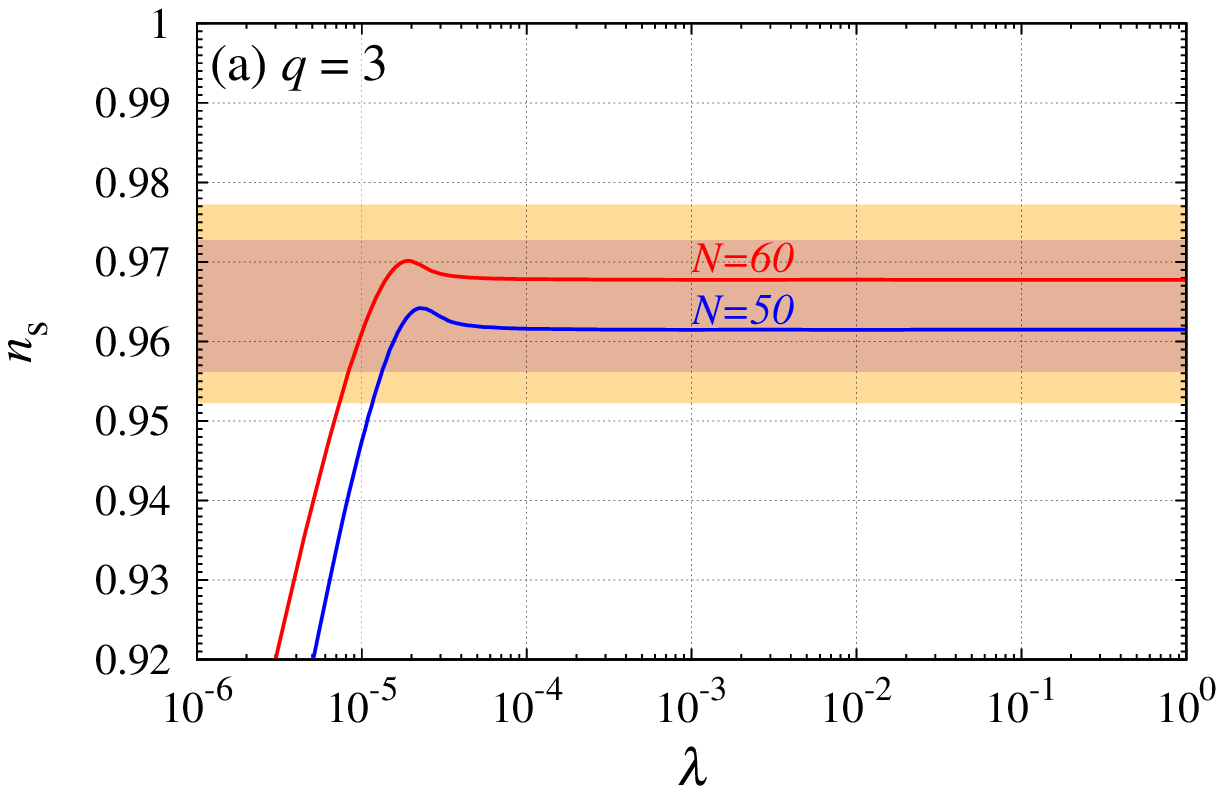} &
\includegraphics[width=80mm]{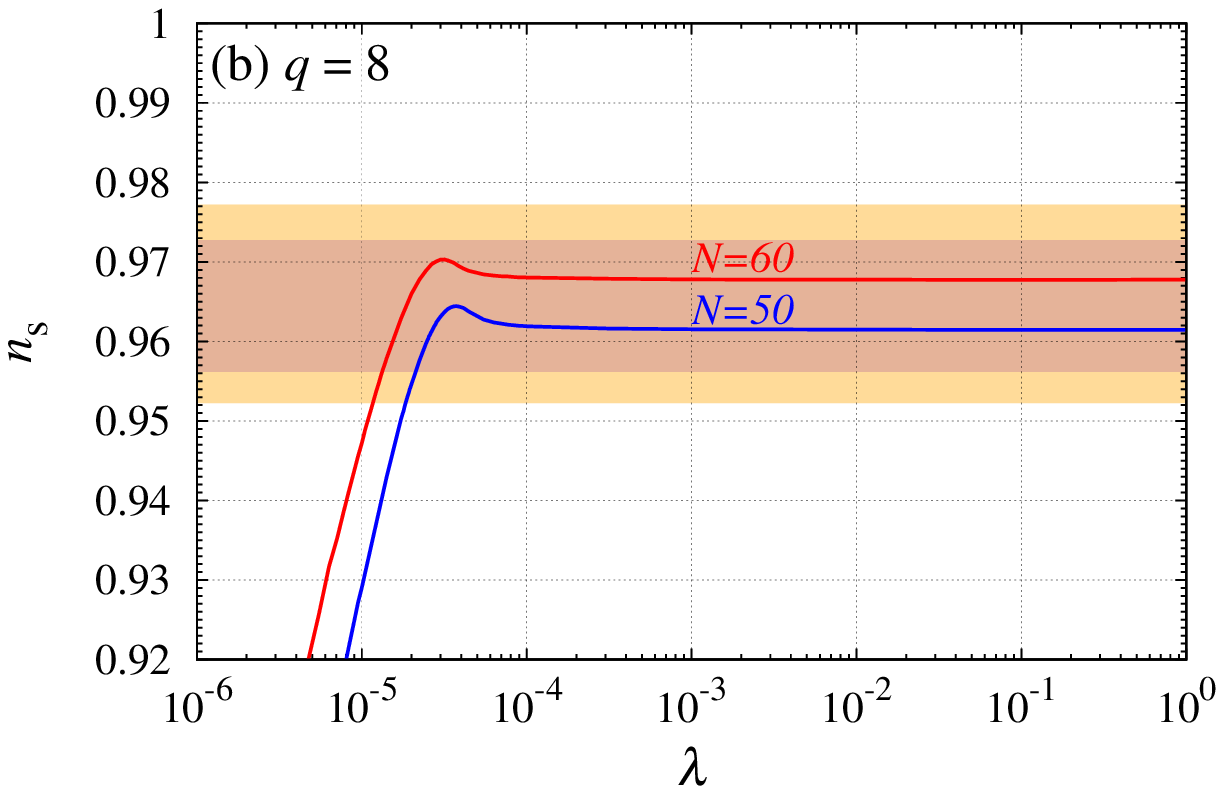}
\end{tabular}
\caption{Spectral index $n_s$ for various $\lambda$ for (a) $q=3$ and (b) $q=8$.}
\label{fig2}
\end{center}
\end{figure}

\begin{figure}[h!]
\begin{center}
\begin{tabular}{cc}
\includegraphics[width=80mm]{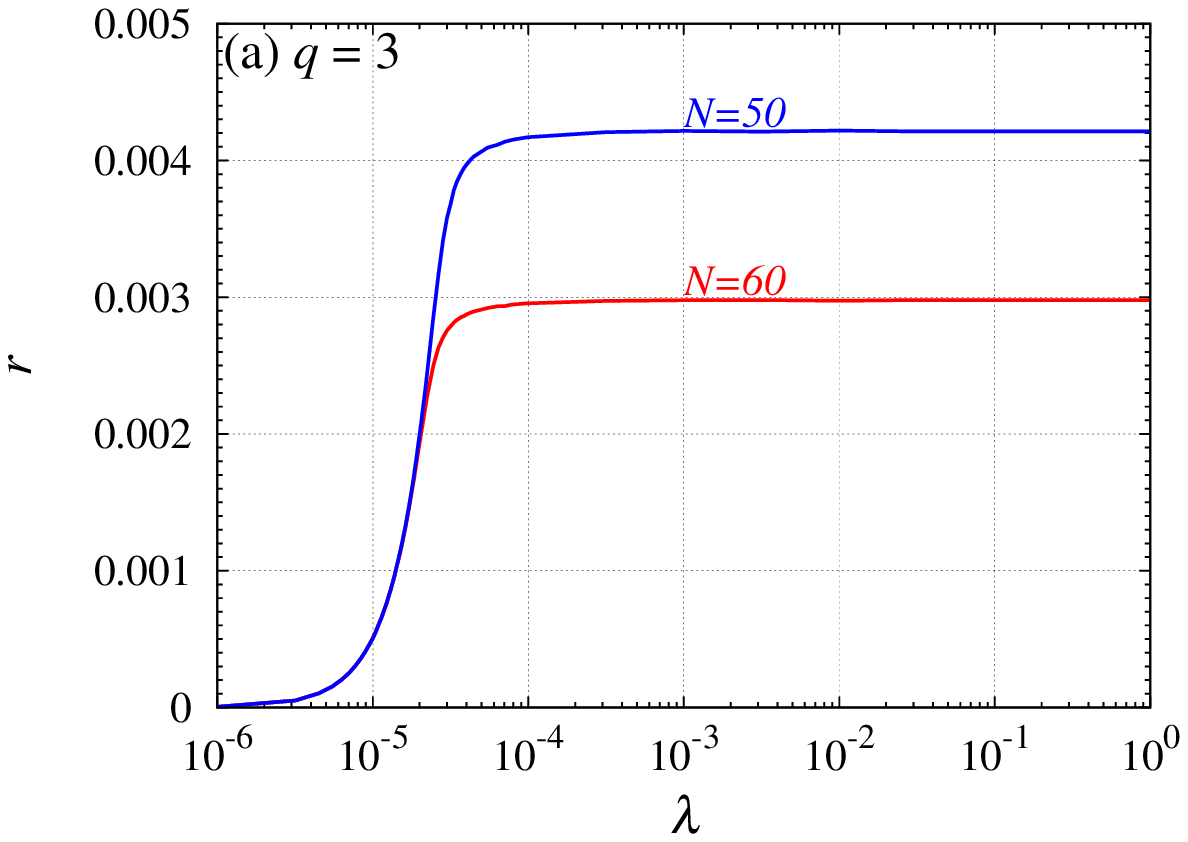} &
\includegraphics[width=80mm]{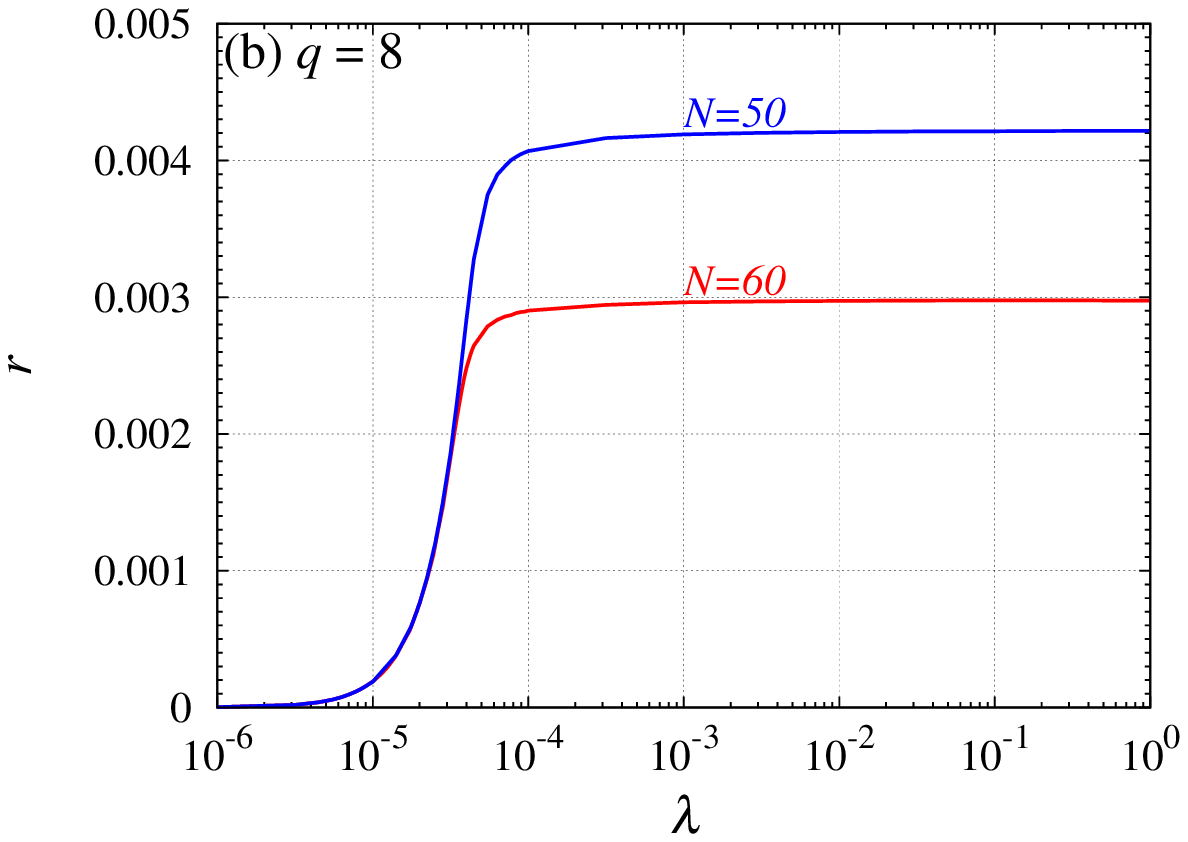}
\end{tabular}
\caption{Tensor-to-scalar ratio $r$ for various $\lambda$ for (a) $q=3$ and (b) $q=8$.}
\label{fig3}
\end{center}
\end{figure}

In order to understand how $n_s$ and $r$ change in the presence of another field with large amplitude, 
we show the amplitude of the inflaton $I$ at $N$=0, 20, 40, and 60 for $q=8$ in Fig. 4. The evolution of the 
inflaton seems changed drastically for $\lambda \lesssim 10^{-4}$. In particular, the excursion of the inflaton gets 
shorter, because $I_{N=60}$ becomes smaller while $I_{\rm end}$ remains almost unchanged. Moreover, 
most of the $e$-folds occur in shorter excursions of the inflaton as $\lambda$ gets smaller. Note that the 
same behavior can be seen for any $q$. We can explain how this happens by considering the potential. 
The inflaton potential at large amplitude is suppressed by the effect of the other field with large vacuum 
expectation values, and the potential becomes flatter. Thus, the tensor-to-scalar ratio $r$ decreases as 
$\lambda$ gets smaller. Eventually, the local maximum appears as $\lambda$ becomes smaller than 
$\sim 10^{-5}$, and most of the $e$-folds occur at the vicinity of the local maximum, as seen in Fig. 4. 
The potential of the Starobinsky model $V_{\rm S}$ and $V(I)$ in Eq.~(9) for $\lambda = 10^{-6}$ and 
$q=8$ are displayed in Fig. 5. There is a local maximum in $V(I)$ for $\lambda = 10^{-6}$ and $q=8$, 
but not in $V_{\rm S}$. Once the local maximum develops as $\lambda$ gets smaller, the curvature 
of the potential becomes larger, and hence the spectral index $n_s$ deviates considerably from that 
of the pure Starobinsky model and comes out of the range allowed by the Planck result.

\begin{figure}[h!]
\begin{center}
\includegraphics[width=80mm]{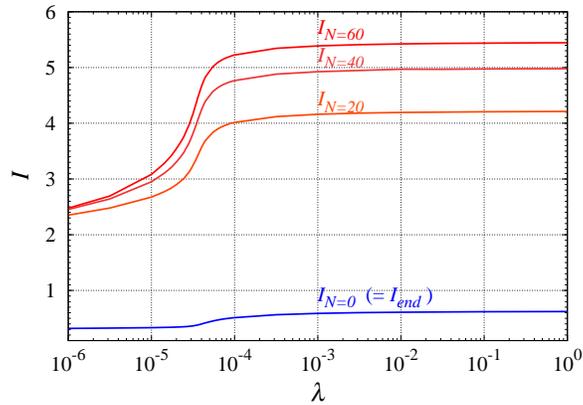} 
\caption{Amplitude of the inflaton $I$ at $N$=0, 20, 40, and 60 for $q=8$.}
\label{fig4}
\end{center}
\end{figure}

\begin{figure}[h!]
\begin{center}
\begin{tabular}{cc}
\includegraphics[width=80mm]{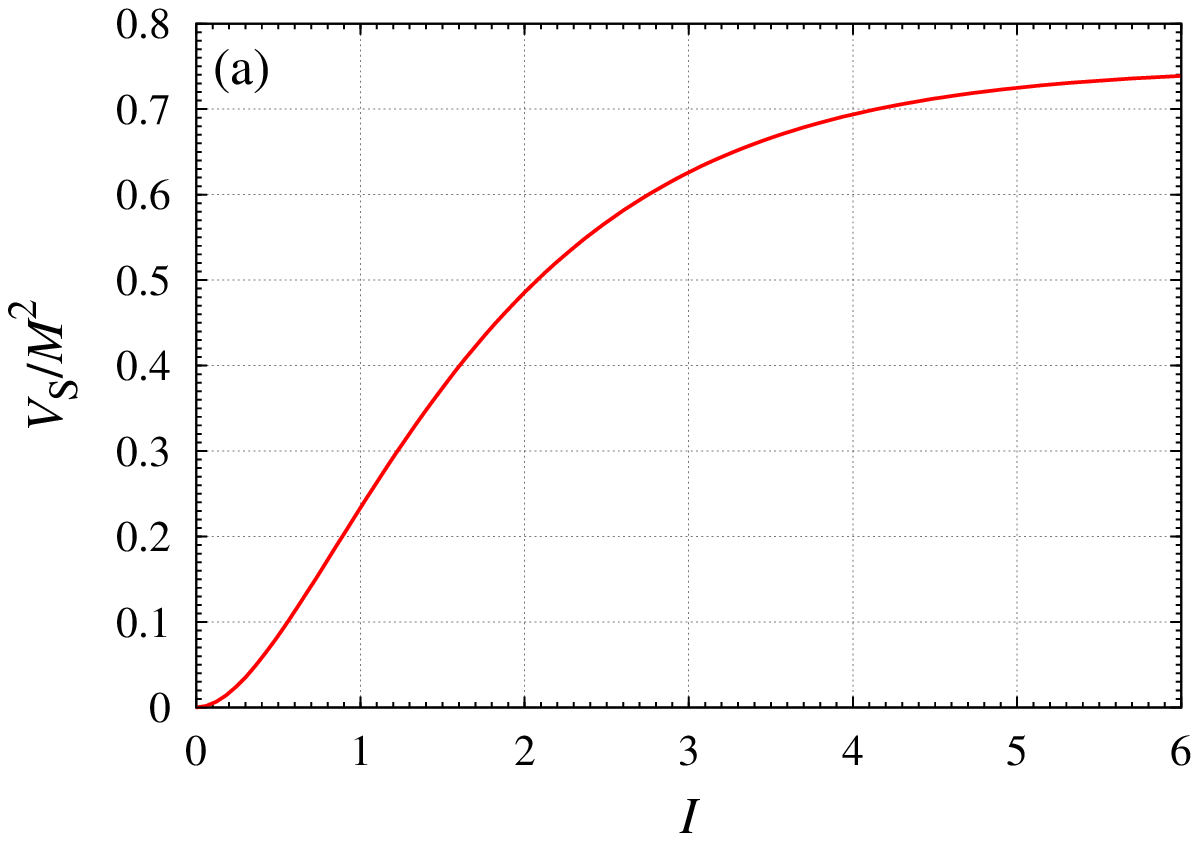}  & 
\includegraphics[width=80mm]{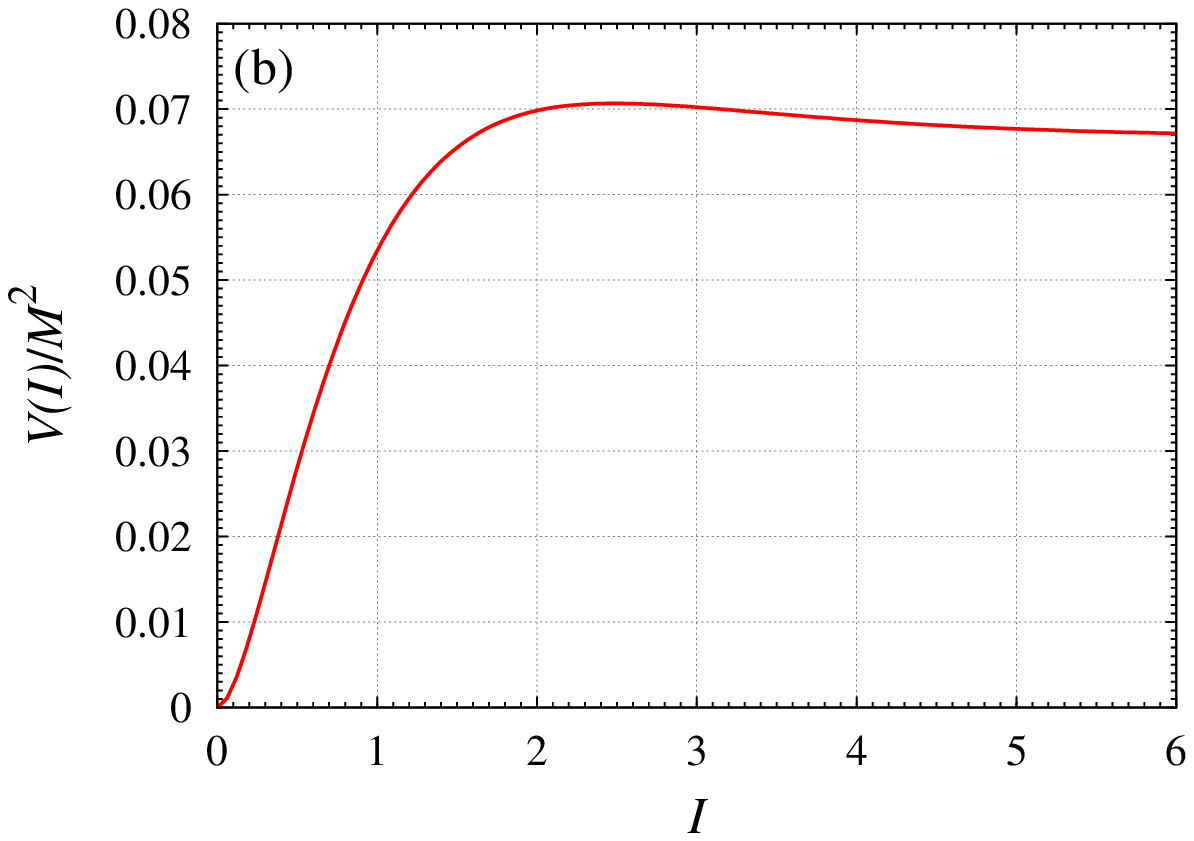}
\end{tabular}
\caption{Potential of the inflaton (a) for the Starobinsky model and (b) in the presence of another field for 
$\lambda = 10^{-6}$ and $q=8$.}
\label{fig5}
\end{center}
\end{figure}

Current CMB observations such as the Planck experiment cannot tell if the Starobinsky inflation 
is affected by the other field with a large amplitude for $\lambda \gtrsim 10^{-5}$. Future experiments may, 
however, reveal such effects of other fields for $\lambda=10^{-5}$--$10^{-4}$. The crucial steps toward this end
are removals of the foreground and the lensing effect, and the forecasts on the capabilities of 
future CMB experiments to constrain the cosmological parameters are reported in Refs.~\cite{CMB1,CMB2,CMB3}.
In particular, Ref.~\cite{CMB3} divides future experiments into two categories, pre- and post-2020, and obtains
the accuracies of the cosmological parameters to be determined as $\sigma(n_s) \sim 2.2\times 10^{-3}$ and
$\sigma(r) \sim 3 \times 10^{-3}$ for the pre-2020 experiments, and $\sigma(n_s) \sim 1.8\times 10^{-3}$ and
$\sigma(r) \sim 1.3 \times 10^{-4}$ for the post-2020 experiments. It seems difficult to distinguish between 
the Starobinsky inflation with and without the effects of other fields by the pre-2020 experiments, but the 
post-2020 experiments may have the ability to confirm the existence of other fields with large amplitudes for 
$\lambda=10^{-5}$--$10^{-4}$, especially due to the very accurate $r$ determination.

\section{Staronbinsky model in supergravity and the negative Hubble-induced mass term for other fields}
The Starobinsky model can be realized in supergravity. For example, it can be derived in superconformal 
theory, where the K\"ahler potential and superpotential are written as~\cite{superconformal}
\begin{equation}
K = S {\bar S} - \frac{\left( \chi - {\bar \chi} \right)^2}{2} - \zeta \left( S {\bar S} \right)^2,
\qquad
W = \frac{M \sqrt{3}}{2} S \left( 1 - e^{- \frac{2\chi}{\sqrt{3}}} \right),
\end{equation}
respectively, where $S$ is the Goldstino superfield, $\chi$ is the scalar field whose real part is 
the inflaton $I$, and $\zeta$ is a constant. Another example utilizes no-scale supergravity, where 
the K\"ahler potential and superpotential are, respectively, given by~\cite{no-scale}
\begin{equation}
K = -3 \log \left( T+ {\bar T} - \frac{|\chi|^{2}}{3} \right),
\qquad
W = \frac{\hat \mu}{2} \chi^2 - \frac{\hat \lambda}{3} \chi^3,
\end{equation}
where $T$ and $\chi$ are complex scalar fields and $\hat \mu$ and $\hat \lambda$ are constants. 
The real part of $\chi$ can be regarded as the inflaton with the moduli $T$ being stabilized~\cite{no-scale}.
 
On the other hand, many light scalar fields may obtain large vacuum expectation values during 
inflation due to a negative Hubble-induced mass in supergravity~\cite{Dine}, and Eq.~(7) is naturally obtained.
It can be achieved by a higher-order term in K\"ahler potential such as 
$\Delta K = a S {\bar S} \Phi {\bar \Phi}$ for superconformal theory or 
$\Delta K = a \chi {\bar \chi} \Phi {\bar \Phi}$ for the no-scale model, where $a$($ >$1) is 
a coupling constant, and $\Phi$ is a complex scalar field whose real part is $\phi$. 
Together with $\Delta W = \frac{\lambda}{p} \Phi^{p}$, we obtain Eq.~(7), with $c$ being $O$(1).

Notice that, in no-scale supergravity, a negative Hubble-induced mass term with $c \sim 0.1$ can 
be realized by one-loop correction in the absence of the above higher-order term in K\"ahler 
potential~\cite{preserving}. In this case, there is less effect of other scalar fields with large amplitude 
on the inflaton potential and dynamics.

\section{Conclusion}
We have considered how the inflaton dynamics in the Starobinsky inflation will change because of 
another scalar field with large amplitude which is naturally obtained due to a negative Hubble-induced 
mass. This effect suppresses the inflaton potential at large amplitudes, and we have found that the 
dynamics of the inflaton will be changed if the amplitude of such other field becomes
close to the Planck scale, $\phi_{*}(I_{N}) \gtrsim 0.7~(0.8)$ for $q=3~(8)$, or, in other words, $\lambda \lesssim 10^{-4}$. For
$\lambda \lesssim 10^{-5}$, which is $\phi_*(I_N) \gtrsim 0.99$, the inflaton potential is distorted too much,
and it is no longer allowed by the current Planck observation. 

We focus on the case for $\lambda \gtrsim 10^{-5}$, where it is allowed at present. In that region, although the
scalar spectral index $n_s$ does not change so much, the tensor-to-scalar ratio $r$ quickly decreases
for $\lambda \gtrsim 10^{-4}$. Since the post-2020 CMB experiments may reach 
$\sigma(r) \sim 10^{-4}$~\cite{CMB1,CMB2,CMB3},
we may have a chance to confirm the presence of such other fields with large amplitude.

Finally, we comment on the case in which the other scalar field is a flat direction in supersymmetry, 
which may result in the Affleck-Dine baryogenesis~\cite{Dine,AD}. Our result seems to show that 
the baryon asymmetry can be created almost maximally without affecting the dynamics of the 
Starobinsky inflation.

\section*{Acknowledgments}
The authors are grateful to Jiro Arafune for helpful conversations.



\end{document}